\documentclass[12pt,a4paper]{article} 
\usepackage{graphicx}

\newcommand{\FC}{F}
\newcommand{\KC}{K}
\newcommand{\nmean}{\langle n \rangle}
\newcommand{\F}{{\cal F}}
\newcommand{\K}{{\cal K}}
\newcommand{\ErrorFC}{\triangle\FC}
\newcommand{\ErrorP}{\triangle P}
\newcommand{\ErrorH}{\triangle H}
\newcommand{\ippi}{IPPI}
\newcommand{\mMean}[1]{\langle m #1 \rangle}
\newcommand{\Errorm}{\triangle m}

\begin{document} 
\begin{center}
          {\large \bf Soft multiple parton interactions as seen in multiplicity
          distributions at Tevatron and LHC} 

\vspace{0.5cm}                   
{\bf I.M. Dremin and V.A. Nechitailo}

\vspace{0.5cm}              
          Lebedev Physical Institute, Moscow 119991, Russia

\end{center}

\begin{abstract}
We analyse the multiplicity distributions of charged particles at Tevatron 
($p\bar p$) and LHC ($pp$) energies in the framework of the independent pair 
parton interactions (IPPI) model. It is shown that the number of soft pair parton 
interactions (and therefore the density of the partonic medium) is large and 
increases with energy. The mean multiplicity at each parton interaction grows 
also with energy. This growth depends on the width of the rapidity window.
The similar conclusions are obtained in the multiladder exchange model (QGSM).
\end{abstract}

\section{Introduction}

Following Feynman, two colliding strongly-interacting high-energy objects
(protons, nuclei etc) are considered as sets of pointlike partons. Parton
interactions determine the outcome of collisions with many particles produced.
They can be either soft, with small transferred momenta, or hard, at large $p_T$.
Most secondary particles are created in soft processes. Usually their
characteristics are described within some phenomenological models with their
Monte Carlo (MC) implementations. Particles (or jets) with high transverse 
momentum are more rare and their production is described by perturbative QCD
albeit at additional assumptions.

One of the general and widely debated (for reviews see \cite{ddk, dgar, gor})
characteristics reflecting mostly the properties of soft processes
is the multiplicity distribution. At energies of tens of GeV the experimental 
data are well fitted by the negative binomial distribution (NBD) first
introduced in \cite{giov}. However at higher energies this fit by a single NBD
becomes inadequate. A shoulder structure appears at high multiplicities.
It is tempting to ascribe it to multiple interactions of pairs of colliding
partons. This was done in several models at the expense of introducing new
parameters. We have proposed \cite{dnec} the most economic model with minimum
adjustable parameters which we call the independent pair parton interactions
(IPPI) model. In fact, we show that there are only two such parameters.
Those are the maximum number of active parton pairs at a given collision energy
and the average multiplicity of particles produced in the collision of a single
pair. At comparatively low energies one pair is active and it leads to NBD
of produced particles which fits experimental data. To deal with the shoulders
at higher energies it is assumed that the number of active parton pairs from 
colliding particles increases with energy. This is in accordance with increase 
of gluon densities at low shares of momentum $x$. Interaction of each additional
pair results in NBD for its products with the same parameters as for a single
pair since the interactions are independent. One is actually tempted to
recognize that this assumption could be valid at asymptotically high energies
where energy conservation becomes unimportant and the number of colliding
partons is infinitely large. We'll try to use it at energies of Tevatron and 
LHC.

It is the well known property of NBD easily seen from its generating function 
(see, e.g., \cite{dgar}) that the convolutions of NBDs again lead to this 
distribution (see \cite{dnec}). NBD is characterized by
two parameters - mean multiplicity $m$ and dispersion $k$. With a maximum number of 
active pairs ($j_{max}$) at a given energy added, we are left with these three 
values. However, we show that one of these parameters, namely the dispersion, 
can be excluded due to some special property of the IPPI model. Thus, only two 
of them should be used for fits of particle multiplicity distributions at any 
energy. To compare, there are seven parameters (including $j_{max}$) in the 
multiladder exchange or quark-gluon string model (QGSM) \cite{kaid} used in 
\cite{mwal} which are to be fitted by additional energy dependent sets of 
experimental data on total and elastic scattering cross sections.
Even the simple fit with two NBDs asks for six parameters to be used if there
are no additional constraints.

Therefore, the IPPI model is the most economic one concerning the adjustable
parameters and can serve as a first approximation for estimates of the global
characteristics of multiple parton interactions. The IPPI model does not imply
that there are no correlations between particles. They are intrinsic in each
binary collision because of assumed NBD  and in their convolution. Surely,
further correlations between these interacting pairs of partons, of both
dynamical and kinematical origin, can be introduced as it is done, for example,
in the multiladder exchange model \cite{kaid} which asks for new parameters. 
Also, there are more rare processes where some pairs interact strongly and
are scattered at large transverse momenta (see,e.g., \cite{snig, berg}). That 
would lead to 2-, 4- and more-jets (or $b\bar b, \gamma $ etc) production. 
However, being interesting by itself, the latter aspect is out of reach of the 
present study.

Let us stress that the knowledge of the number of active parton pairs provides
information about the density of the hadronic matter formed in collisions and
about its evolution with energy. We shall see that this density is high and
grows with energy increase. These findings favor the enlarged role of 
collective effects at LHC energies.

\section{Application of the IPPI model}

The main equation of the IPPI model obtained in \cite{dnec} is
\begin{equation}\
P(n;m,k)=\sum _{j=1}^{j_{max}}w_jP_{NBD}(n;jm,jk).   \label{main}
\end{equation}
It states that the probability of the $n$-particle production channel is defined
by the sum of NBDs with shifted maxima ($jm$) and larger widths ($jk$) for 
independent parton 
collisions weighted by their probabilities $w_j$. The total number of particles
$n$ is equal to the sum of particles produced in all active pairs collisions.
At asymptotically high energies the probability for $j$ pairs of independent
interactions $w_j$ is the product of $j$ probabilities for one pair so that the
normalization condition
\begin{equation}
\sum_{j=1}^{j_{max}}w_j=\sum_{j=1}^{j_{max}}w_1^j=1     \label{prob}
\end{equation}
determines $w_1$ if $j_{max}$ is known at a given energy. In fact, the value
of $w_1$ ranges between 1 at low energies (for $j_{max}=1$) and 0.5 at 
asymptotics where $j_{max}$ tends to infinity. Thus, all values $w_j$ are
calculated from Eq. (\ref{prob}) if $j_{max}$ is defined. We show them in Table I
for 6-10 active pairs which happen to be important at TeV energies.

\begin{table}[t]
 \begin{tabular}{|c|c|c|c|c|c|c|}
\hline      &\multicolumn{2}{c|}{IPPI}   &\multicolumn{4}{c|}{QGSM(14TeV)}  \\ \hline
$j_{max}$&     6      &    7       &        7  &     8     &    9       &   10       \\ \hline
$w_1 $   &  0.5041383 &  0.5020171 & 0.4272531 & 0.4237175 & 0.4222109  & 0.4216202  \\ \hline
$w_2 $   &  0.2541554 &  0.2520211 & 0.2238815 & 0.2220288 & 0.2212393  & 0.2209298  \\ \hline
$w_3 $   &  0.1281295 &  0.1265189 & 0.1441260 & 0.1429333 & 0.1424251  & 0.1422259  \\ \hline
$w_4 $   &  0.0645950 &  0.0635146 & 0.0945266 & 0.0937443 & 0.0934110  & 0.0932803  \\ \hline
$w_5 $   &  0.0325648 &  0.0318854 & 0.0588581 & 0.0583711 & 0.0581635  & 0.0580822  \\ \hline
$w_6 $   &  0.0164172 &  0.0160070 & 0.0337454 & 0.0334661 & 0.0333471  & 0.0333005  \\ \hline
$w_7 $   &            &  0.0080358 & 0.0176093 & 0.0174636 & 0.0174015  & 0.0173771 \\  \hline
$w_8 $   &            &            &           & 0.0082753 & 0.0082459  & 0.0082344  \\ \hline
$w_9 $   &            &            &           &           & 0.0035557  & 0.0035507  \\ \hline
$w_{10}$ &            &            &           &           &            & 0.0013988 \\ \hline
\end{tabular}
\end{table}
%        4 &      5    &    8      &     9       &    10        &     4    &     5     &      6    
% 0.5187901&  0.5086604& 0.5009942 &  0.5004931  &  0.5002455   &0.4801745 & 0.4503824 & 0.4349116 
% 0.2691431&  0.2587354& 0.2509952 &  0.2504934  &  0.2502455   &0.2516124 & 0.2360012 & 0.2278945 
% 0.1396288&  0.1316084& 0.1257471 &  0.1253702  &  0.1251842   &0.1619781 & 0.1519283 & 0.1467095 
% 0.0724380&  0.0669440& 0.0629986 &  0.0627469  &  0.0626228   &0.1062350 & 0.0996437 & 0.0962210 
%          &  0.0340518& 0.0315619 &  0.0314044  &  0.0313268   &          & 0.0620444 & 0.0599132 
%          &           & 0.0158123 &  0.0157177  &  0.0156711   &          &           & 0.0343503 
%          &           & 0.0079219 &  0.0078666  &  0.0078394   &          &           &           
%          &           & 0.0039688 &  0.0039372  &  0.0039216   &          &           &           
%          &           &           &  0.0019705  &  0.0019618   &          &           &           
%          &           &           &             &  0.0009814   &          &           &           

The IPPI model predicts new special features of moments of the multiplicity
distribution which impose some constraints on the parameters and allow to
get rid of one of them, namely $k$. The factorial moments of the distribution
(\ref{main}) are
\begin{equation}
F_q=\sum_nP(n)n(n-1)...(n-q+1)=\sum_{j=1}^{j_{max}}w_j\frac {\Gamma (jk+q)}
{\Gamma (jk)} \left (\frac {m}{k}\right )^q=f_q(k)\left (\frac {m}{k}\right )^q
\label{Fq}
\end{equation}
with
\begin{equation}
f_q(k)=\sum_{j=1}^{j_{max}}w_j\frac {\Gamma (jk+q)}{\Gamma (jk)}=
k\sum_{j=1}^{j_{max}}w_jj(jk+1)...(jk+q-1).       \label{fq}
\end{equation}
Herefrom one gets the relation
\begin{equation}
m=k\left (\frac {F_q}{f_q(k)}\right )^{1/q}.     \label{frac}
\end{equation}
This relation states that the righthand side with the definite ratio of
$q$-dependent functions should be independent of $q$ for some value of $k$.
It opens the way to the combined fit of experimental mutliplicity distributions
with the requirement of $q$-independence of $m$ (see Appendix 1).

This requirement is very strong and can be satisfied at some special values
of $w_j$ unknown to us. It would be too naive to expect it to be precisely 
satisfied in a simplified model. Nevertheless, we try to find such fits of
experimental multiplicity distributions by the IPPI model which minimize the
decline of $m$ as a function of $q$ from constancy. After doing this,
we get the important information about the maximum number of active parton
pairs at a given energy and its evolution with energy. It provides the clear 
insight into the dynamics of soft interactions and properties of the
hadronic medium formed during the collision.

Beside the factorial moments $F_q$ (\ref{Fq}) we use the cumulants
\begin{equation}
K_q=F_q-\sum _{r=1}^{q-1}\frac {(q-1)!}{r!(q-r-1)!}K_{q-r}F_r
\label{Kq}
\end{equation}
and their ratios
\begin{equation}
H_q=K_q/F_q
\label{Hq}
\end{equation}
which possess some specific oscillating behaviour (see \cite{dgar, drem, cdgt}).
This is the complementary (and sometimes more sensitive!) approach. 

We have used both direct fits of multiplicity distributions and the fits of
$H_q$ moments at energies of Tevatron 1.8 TeV (for $p\bar p$ interactions) 
\cite{E735} and LHC 0.9, 2.36 and 7 TeV (for $pp$ interactions) \cite{CMS}. 
The detailed description of our procedure is presented in Appendix 1. The 
extrapolation of the obtained results admits predictions at 14 TeV which we 
also show below.

\begin{figure}
%fitPmq1800.gnu
 \includegraphics[width=\textwidth]{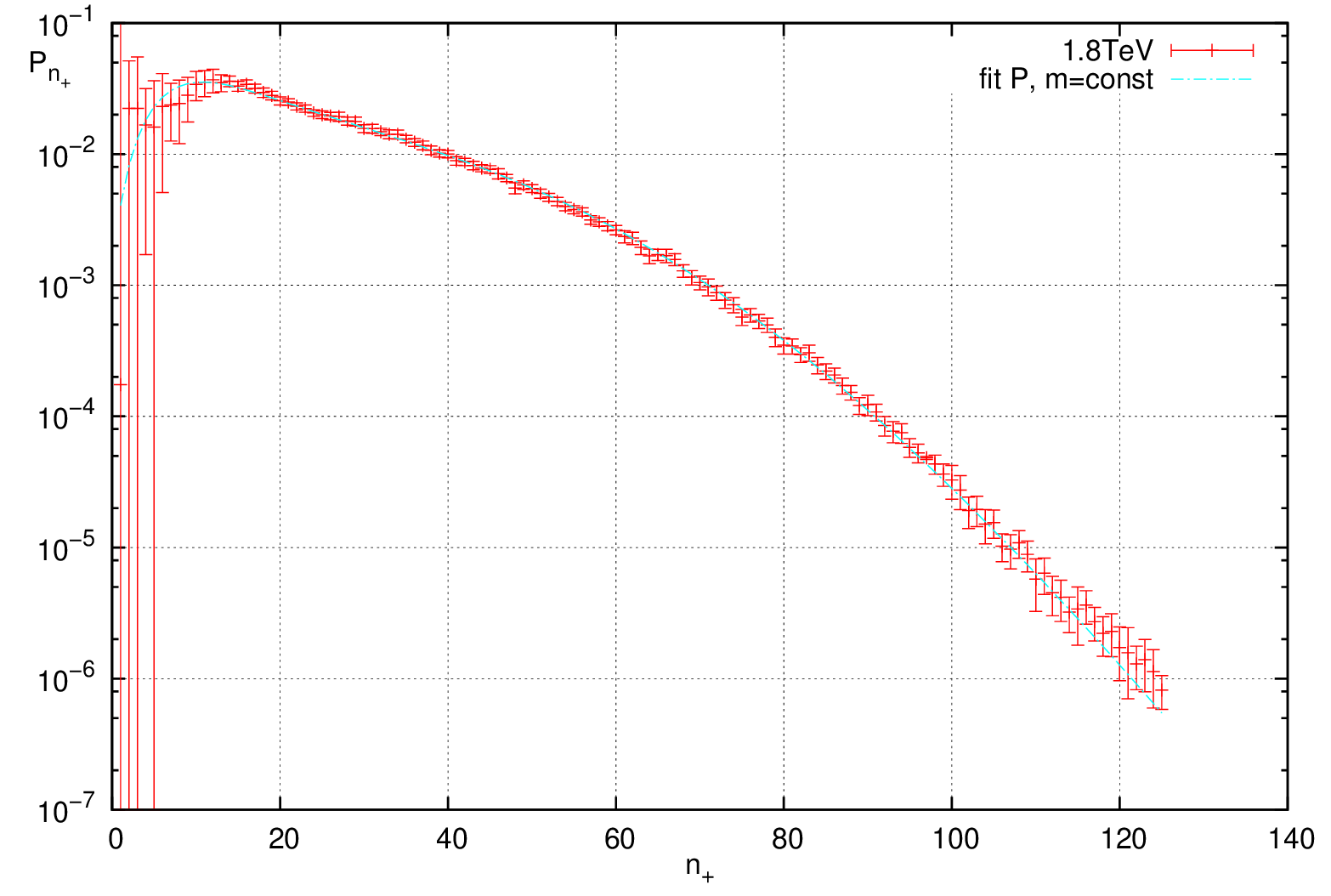}
 \caption{}
 \label{fitPmq1800}
\end{figure}

\begin{figure}
%fitHmq1800.gnu
 \includegraphics[width=\textwidth]{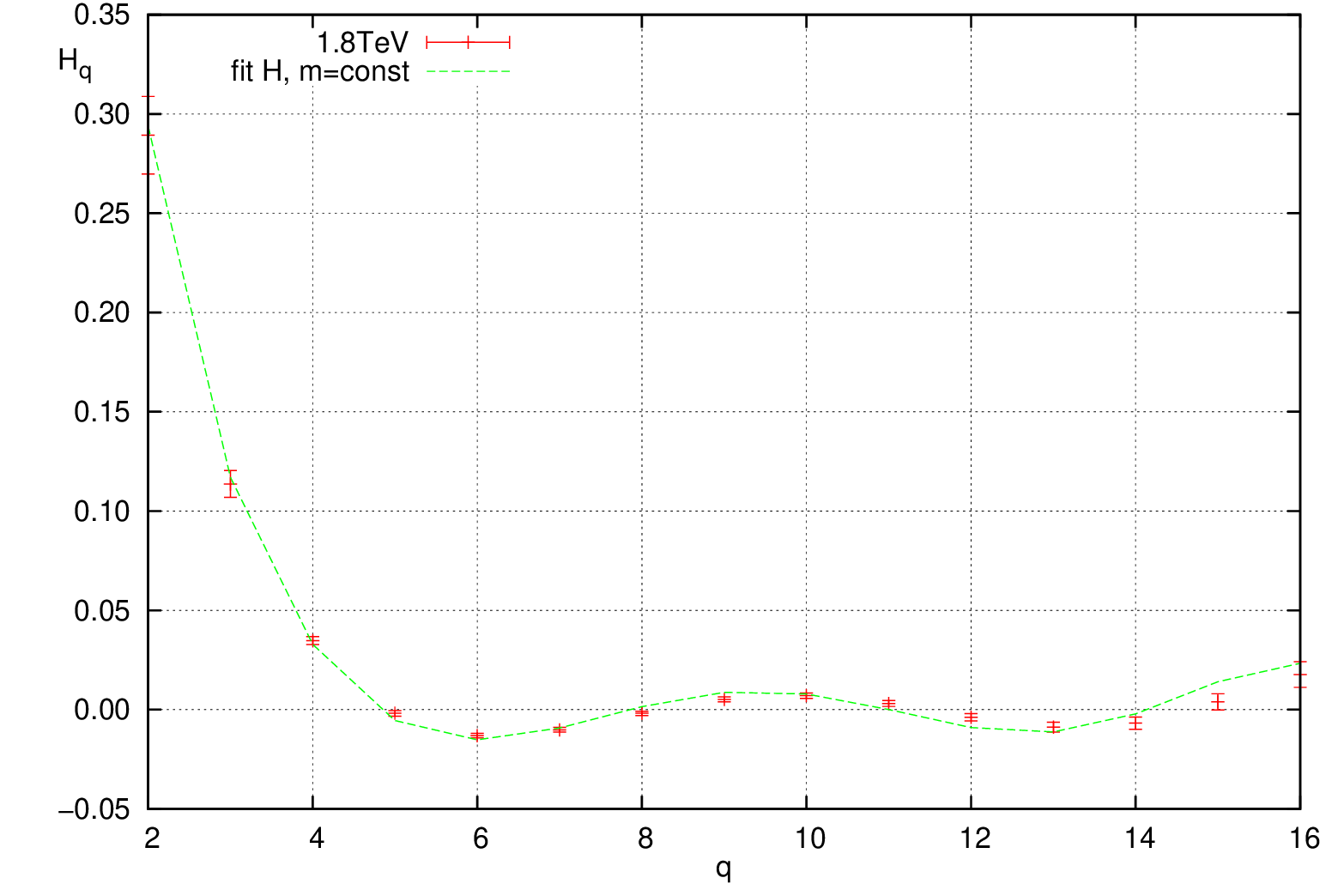}
 \caption{}
 \label{fitHmq1800}
\end{figure}

\begin{figure}
%m_q.gnu
 \includegraphics[width=\textwidth]{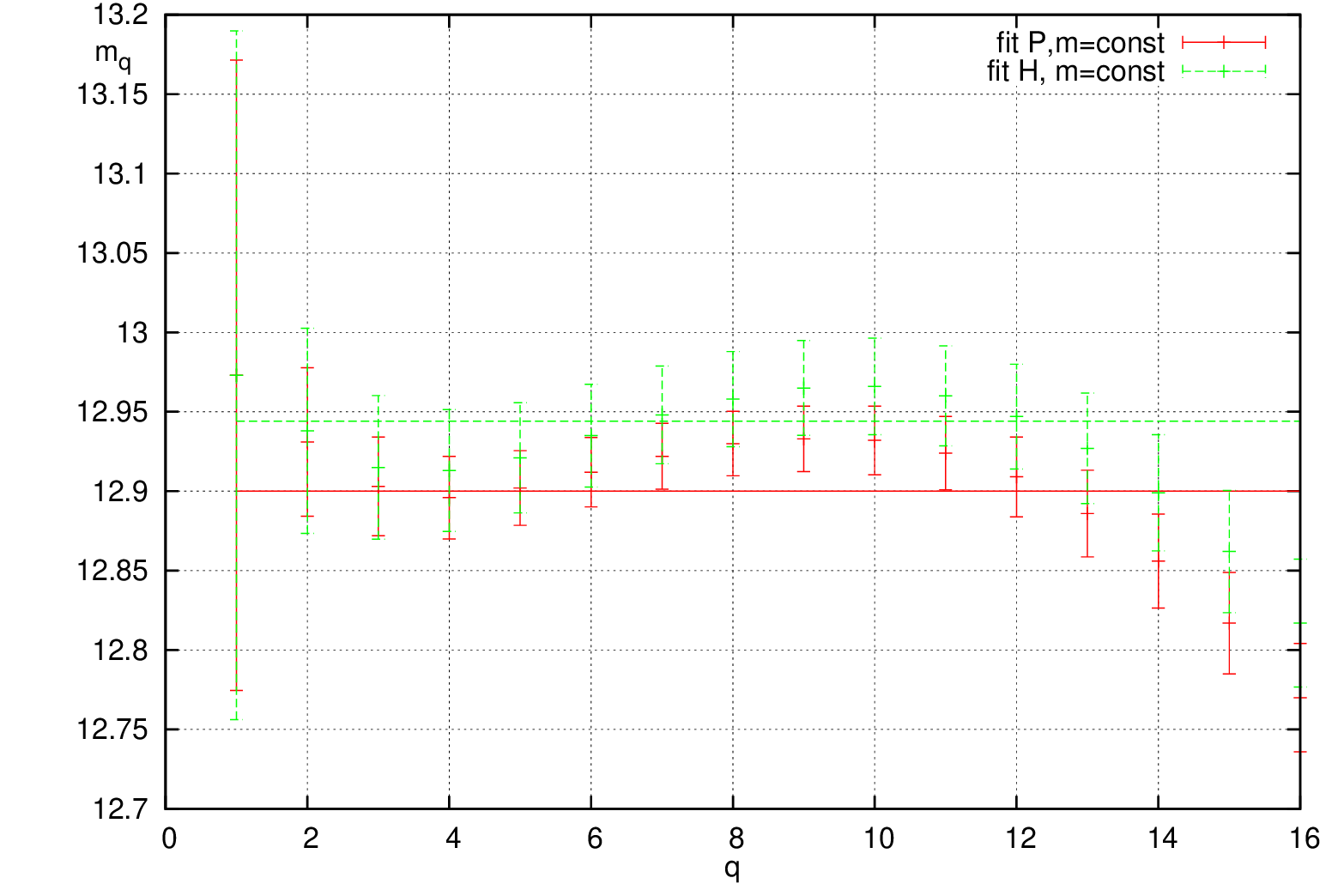}
 \caption{}
 \label{fit_mq1800}
\end{figure}

The fit of the multiplicity distribution at 1.8 TeV was first done in our paper
\cite{dnec}. Now using the improved procedure described in Appendix 1 we
confirmed the stability of previous fits and their parameters as seen from good 
fit in Fig.~\ref{fitPmq1800}. The values of fit parameters are $k=4.42$ and 
$m=12.944 \pm 0.04$ which practically coincide with those in \cite{dnec}.
The corresponding fit of $H_q$ is presented in Fig.~\ref{fitHmq1800} ($k=4.36$, 
$m=12.90 \pm 0.05$). The values of $k$ and $m$ are close in both approaches. 
The independence of $m$ on $q$ is satisfied within the limits less than 1 $\% $ 
as seen in Fig.~\ref{fit_mq1800}. Within these limits the value of $k$ does
not play a role and can be considered fixed by constancy of $m$ at a given
energy. 

The maximum number of the 
pair parton interactions is $j_{max}$(1.8 TeV)$\approx $4. 

The same procedure was used for the LHC $pp$-data at 0.9, 2.36 and 7 TeV. 
The experimental distributions are available for limited pseudorapidity
intervals. Namely they have been used by us in distinction to the above fit
at Tevatron energy where the data were given for the total pseudorapidity
window with some extrapolation to the fragmentation region. Beside some
arbitrariness of such extrapolation, the multiplicity distribution for the
full phase space is influenced by energy-momentum and charge conservation.
These constraints are less important in restricted domains. Therefore, the
distributions can be expected to be more sensitive to the underlying dynamics. 

\begin{figure}
%m_all.gnu
 \includegraphics[width=\textwidth]{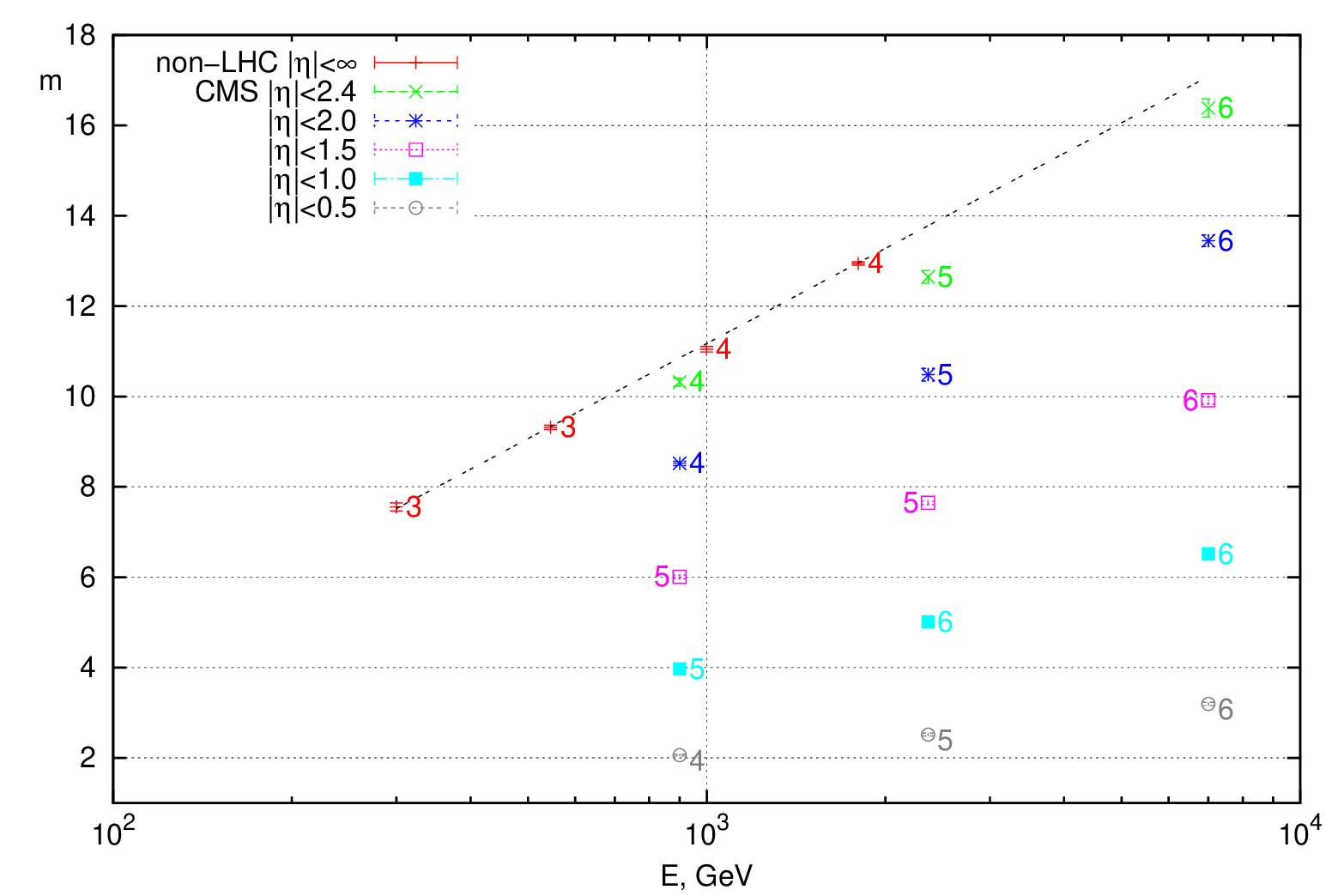}
 \caption{}
 \label{m_all}
\end{figure}

The parameter
$m$ shows now what part of the multiplicity produced in a single pair parton
interaction reaches the analysed pseudorapidity interval. Since intervals
$\vert \eta \vert <$2.4 cover the region of almost flat pseudorapidity
distribution one would expect that the ratios of corresponding values of $m$
are approximately equal to the ratios of the intervals themselves. On the 
contrary, in the fragmentation region $\vert \eta \vert >$2.4 the
distribution drops down. Thus the values of $m$ for 2.4 must be close to
the parameters obtained from extrapolated Tevatron data. All these features
as well as the explicit energy dependence of $m$ are clearly seen in Fig.~\ref{m_all} 
where some results for lower energies from our paper \cite{dnec} are also shown.

In the same Figure we plot the numbers of active parton pairs from our fits
for each pseudorapidity interval near corresponding points. They are quite 
stable for a fixed energy and rise from 4 at 0.9 TeV to 6 at 7 TeV.
The corresponding contributions to the total mean multiplicity (equal to 31.4
at 7 TeV for the interval $\vert \eta \vert <$2.4) of these 6 pairs 
interactions are 8.3+8.3+6.3+4.2+2.7+1.6.

% IPPI
%    w_j         w_j*<n_j>
% 1 0.5041383  8.2595862
% 2 0.2541554  8.3279433
% 3 0.1281295  6.2975723
% 4 0.0645950  4.2323781
% 5 0.0325648  2.6634164
% 6 0.0164172  1.6001585
%exp=30.4303645 fit=31.3810549
% QGSM
% 1 0.4356620  5.9096435
% 2 0.2261431  6.1351465
% 3 0.1426641  5.8056030
% 4 0.0906092  4.9163197
% 5 0.0541212  3.6703463
% 6 0.0295755  2.4054114
% 7 0.0146507  1.3861726
% 8 0.0065743  0.7037829
%   30.4303645 30.9324258

The achieved good fits of the shapes of multiplicity distributions with 
above parameters assure that the energy dependence of mean multiplicity
and higher moments are also well reproduced.

\begin{figure}
%7TeV.gnu
 \includegraphics[width=\textwidth]{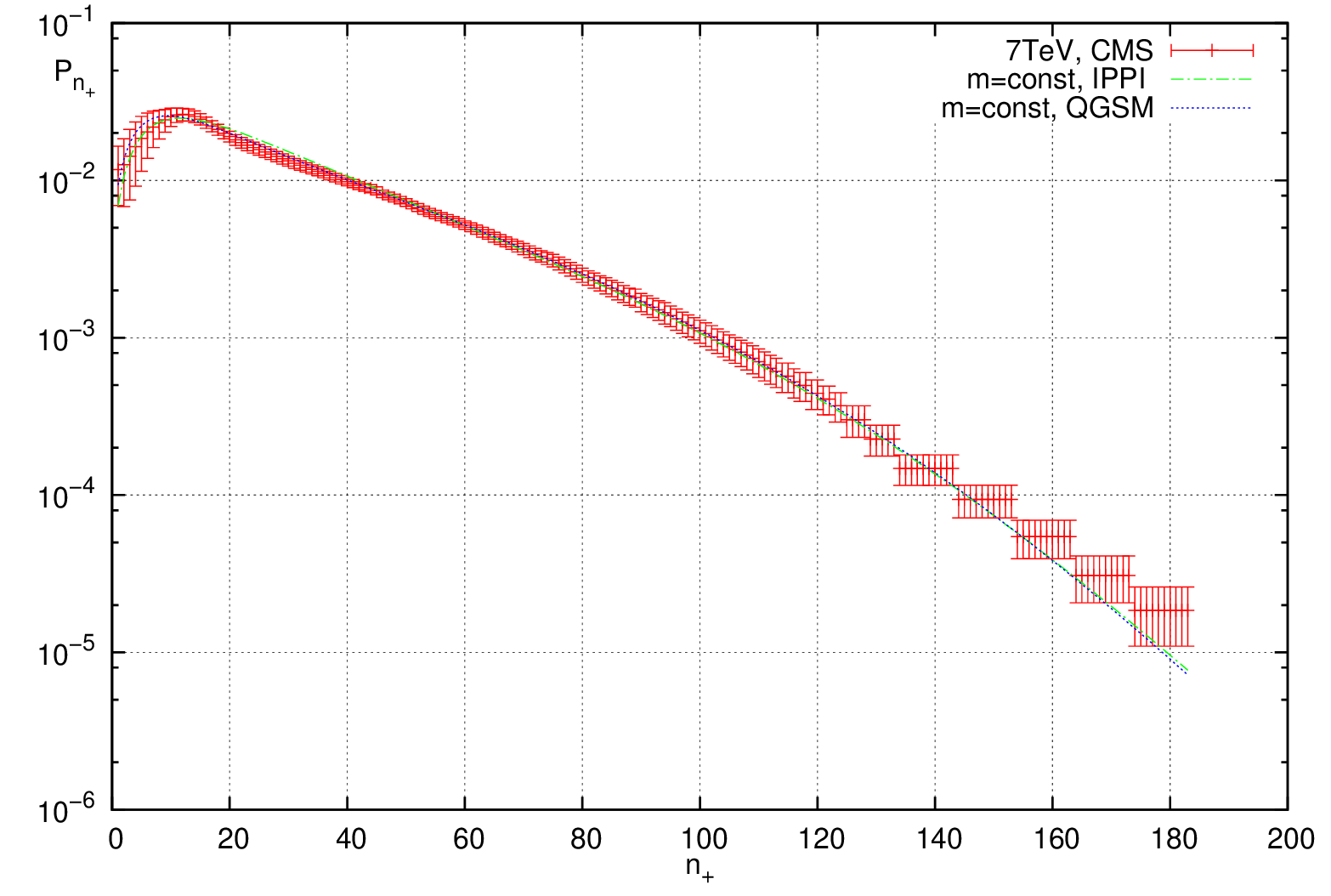}
 \caption{}
 \label{fitPmq7000log}
\end{figure}

We have compared our conclusions with those which one would obtain from the
multiladder exchange model (QGSM) \cite{kaid} used in \cite{mwal}. The lower  
mean multiplicity of a single pair parton interaction and, correspondingly,
somewhat larger numbers of ladders are allowed there due to the
wider spread of probabilities $w_j$. For example, at 7 TeV they would be
13.6 with 8 active pairs of partons. For these fits we have chosen the
values of 7 adjustable parameters exactly equal to those used in \cite{mwal}
and can not say how sensitive to their variations are the results.
The overall fit of the multiplicity distribution for $\vert \eta \vert <$2.4
according to the multiladder model is somewhat better than in the IPPI model
(see Fig.~\ref{fitPmq7000log}). Corresponding $\chi ^2$/dof are 62/127 and 
131/127.

\begin{figure}
%14TeV.gnu
 \includegraphics[width=\textwidth]{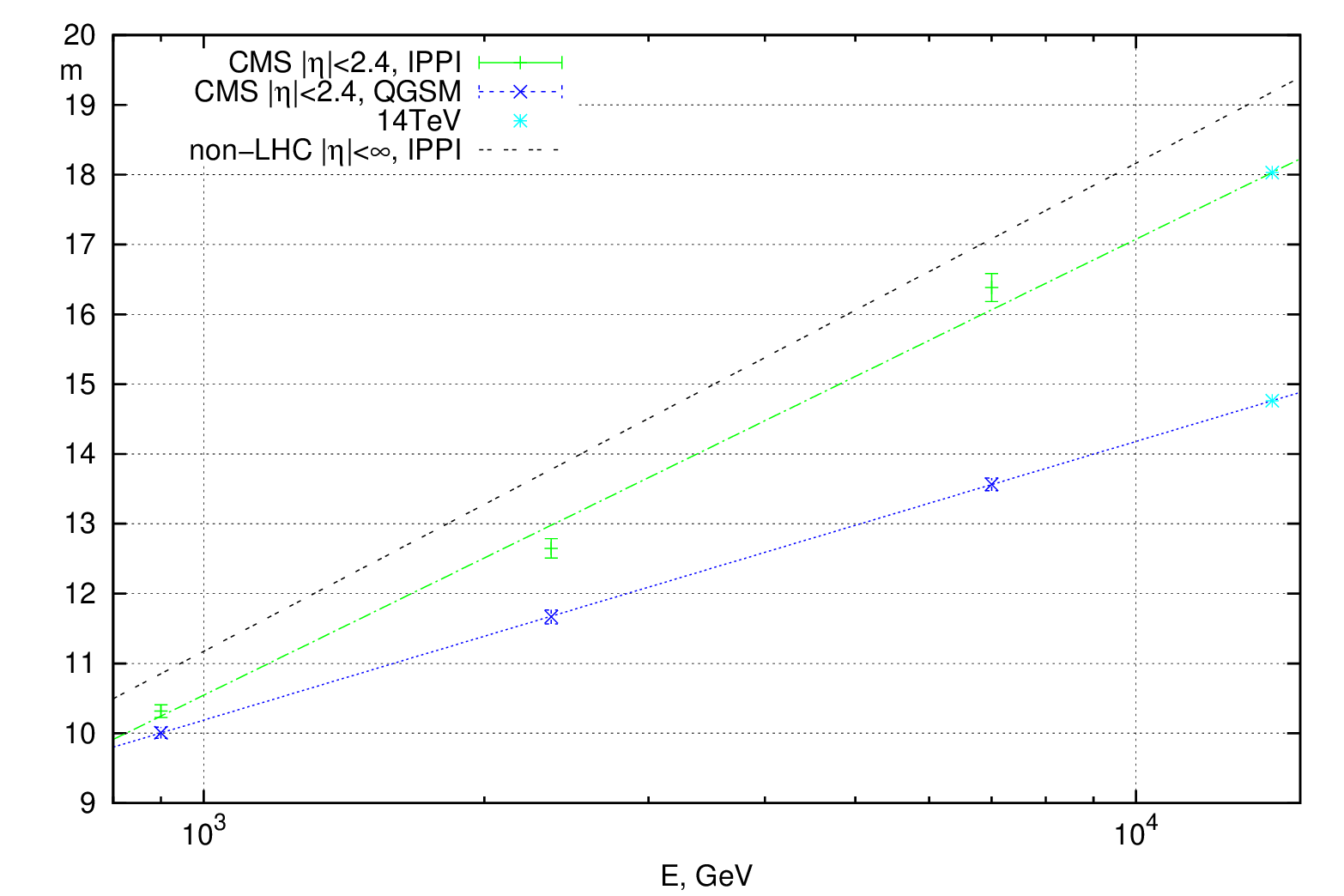}
 \caption{}
 \label{m_2.4CMS}
\end{figure}

\begin{figure}
%j_max.gnu
 \includegraphics[width=\textwidth]{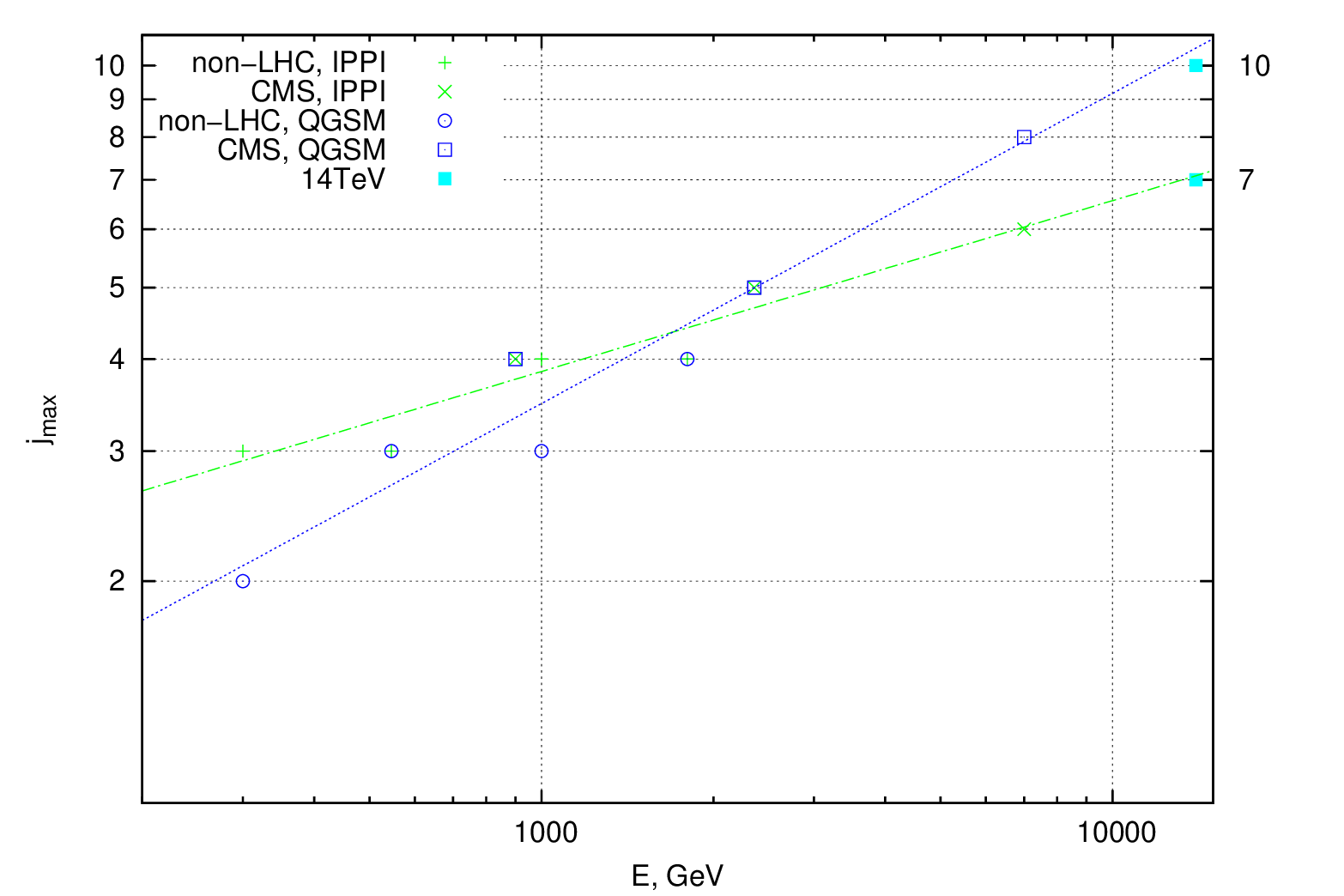}
 \caption{}
 \label{j_max}
\end{figure}

The qualitative conclusions about the energy increase of the mean
multiplicity at a single parton interaction (see Fig.~\ref{m_2.4CMS}) 
and the number of such interactions (see Fig.~\ref{j_max})
are strongly supported in both approaches.

\begin{figure}
%14TeV_Pn.gnu
 \includegraphics[width=\textwidth]{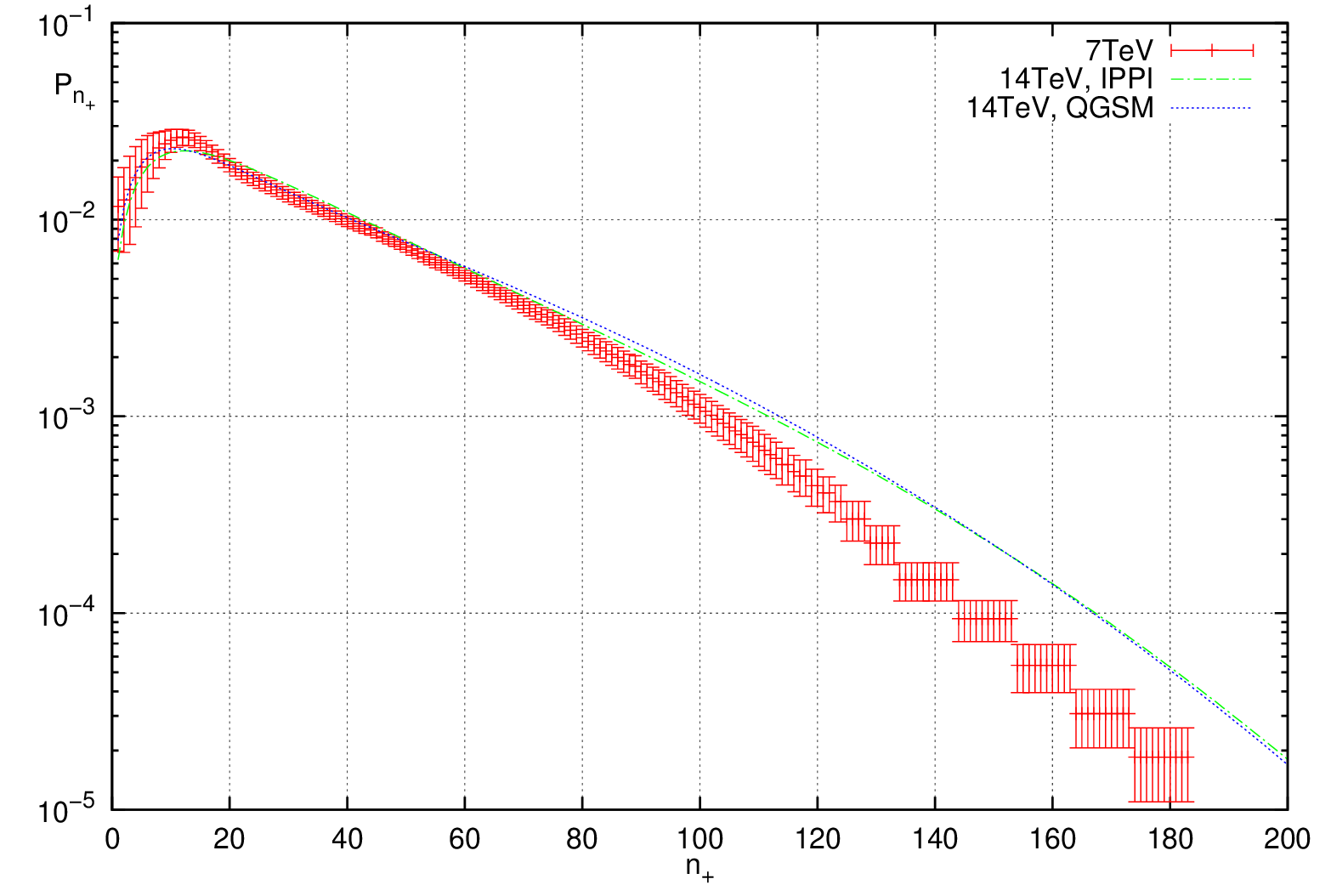}
 \caption{}
 \label{fitPmq14000log}
\end{figure}

In Fig.~\ref{fitPmq14000log} we plot the predictions at 14 TeV for 
$\vert \eta \vert <$2.4 in both models. They are obtained by using the 
extrapolated values of $m$ shown in Fig.~\ref{m_2.4CMS}. The predicted
numbers of active parton pairs at 14 TeV range from 7 in IPPI to 10 in QGSM.
In general, we see that the difference between these models  becomes
noticeable only at energies as high as 7 TeV even though it is still not
very well pronounced in Fig.~\ref{fitPmq14000log}. 

We conclude that the high density partonic medium is formed not only in
heavy-ion collisions but also at high energy $pp$-interactions. 
The number of active parton pairs taking part in soft $pp$ interactions 
increases with energy reaching values 6 - 8 at 7 TeV and could be 7 - 10 
at 14 TeV, i.e their density is higher at higher energies. 
Therefore the theoretical account of multiple parton interactions is 
absolutely necessary at high energies. The mean number of particles created 
at a single interaction of the pair of partons also increases with energy. 
This is quite natural because the structure functions are modified 
correspondingly. The collective effects should become more pronounced
at LHC energies.

\section{Appendix 1}

The experimental data on multiplicity distributions were obtained at Tevatron
(for $p\bar p$) and LHC (for $pp$) in various pseudorapidity intervals. Their
moments can be computed and compared with predictions of the IPPI model.
To get smaller numerical values at the intermediate stages of computing, we 
actually deal with factorial and cumulant moments normalized by $\nmean^q(q-1)!$.
\begin{equation}
\F_q \equiv \frac{\FC_q}{\nmean^q (q-1)!} = \sum_{n=q}P_n \frac{n}{\nmean}
\prod_{r=1}^{q-1} \frac{n/r-1}{\nmean},
\label{F}
\end{equation}
\begin{equation}
 \K_q = \F_q - \sum_{r=1}^{q-1} \frac{\F_r}{r}\K_{q-r}.
\label{KF}
\end{equation}
The values of $H_q$ are not changed by this normalization.

To estimate the errors in moments induced by the experimental error bars of 
$P_n$ one usually assumes that the latter ones are independent for different $n$
and gets
\begin{equation}
(\ErrorFC_q)^2 = \sum_{n}\left(\frac{\partial\FC_q}{\partial P_n}\right)^2 (\ErrorP_n)^2
                    = \sum_{n} (n(n-1)\ldots(n-q+1))^2 (\ErrorP_n)^2,
\label{ErrorFC}
\end{equation}
\begin{equation}
(\ErrorH_q)^2 = \sum_{n}\left(\frac{\FC_q\frac{\partial\KC_q}{\partial P_n}
-\KC_q\frac{\partial\FC_q}{\partial P_n}}{\FC_q^2}\right)^2 (\ErrorP_n)^2,
\label{ErrorH}
\end{equation}
where $\partial K_q/\partial P_n$ is obtained by differentiating (\ref{Kq}).

Moreover, one should include the requirement of $q$-independence of $m$ in the 
fitting procedure minimizing the decline of values given by Eq. (\ref{frac}) 
from a constant when calculations are done within some model. Then one minimizes
the following functions
\begin{equation}
 \min\limits_{k,m}^{} f^{Err}_P(k,m)=
 \min\limits_{k,m}^{}\sum_{n}\left(\frac{ P^{\ippi}_n(k,m) -P_n}{\ErrorP_n} \right)^2
+ \sum_{q=2}^{q_{max}}\left(\frac{m(q,k)-m}{\Errorm(q,k)}\right)^2,
\label{fitPmq=const}
\end{equation}
\begin{equation}
 \min\limits_{k}^{} f^{Err}_H(k)=
 \min\limits_{k}^{}\sum_{q=2}^{q_{max}}\left(\frac{ H^{\ippi}_q(k) -H_q}{\ErrorH_q} \right)^2
+ \left(\frac{m(q,k)-\mMean{(k)}}{\Errorm(q,k)}\right)^2.
\label{fitHmq=const}
\end{equation}
Here $m(q,k)$ is computed according to (\ref{frac}) in the IPPI model.
As $m$ in Eq.~(\ref{fitPmq=const}) is one of the parameters to be minimized 
it does not contain experimental errors (in contrast to $\mMean{(k)}$ in 
Eq.~(\ref{fitHmq=const})) and the denominator in the second sum in 
Eq.~(\ref{fitPmq=const}) can be computed as
\begin{equation}
\Errorm(q,k)=\frac{m(q,k) \ErrorFC_q}{q \FC_q} .
\end{equation}

The case of Eq. (\ref{fitHmq=const}) asks for a more complicated formula which is not 
shown here. Let us note that all the sums are not normalized to the number
of particles and the number of the moments since it enlarges the relative
weight of the constancy of $m$ while our goal is to get the best fit of the
multiplicity distributions at the satisfactory fit of additional conditions.

Using these errors one can find out the parameters of the IPPI model in the
twofold way: either by the direct fit of multiplicity distributions $P_n$,
or by fits of computed values of $H_q$. We show the results of both approaches.
The latter one is more transparent and the error bars are easily visualized.

There is another problem which becomes important. Namely, one must decide
what is the maximum rank (the value of $q_{max}$) to be used in Eq. (\ref{fitPmq=const}).
To answer this question we introduce a simple criterium related to the
experimental error of $\Delta H_q$. We compute the dispersion of $\Delta H_q$
according to the standard expression
\begin{equation}
D^H_q=\langle \ErrorH^2\rangle_q - \langle \ErrorH\rangle^2_q,
\end{equation}
where 
\[
%  mean2_:=mean2_+(sqr(Error_)-mean2_)/Count_;
   \langle \ErrorH^2\rangle_q = \langle \ErrorH^2\rangle_{q-1}+
   \frac{1}{q-1}(\ErrorH^2_q - \langle \ErrorH^2\rangle_{q-1}),
\]
and
\[
%mean_:=mean_+(Error_-mean_)/Count_;
\langle \ErrorH\rangle_q = \langle \ErrorH\rangle_{q-1}+
   \frac{1}{q-1}(\ErrorH_q- \langle \ErrorH\rangle_{q-1})
\]
with $\langle \ErrorH\rangle_{1}=\langle \ErrorH^2\rangle_{1}=0$ by definition.

\begin{figure}[h]
%7TeV.gnu
 \includegraphics[width=\textwidth]{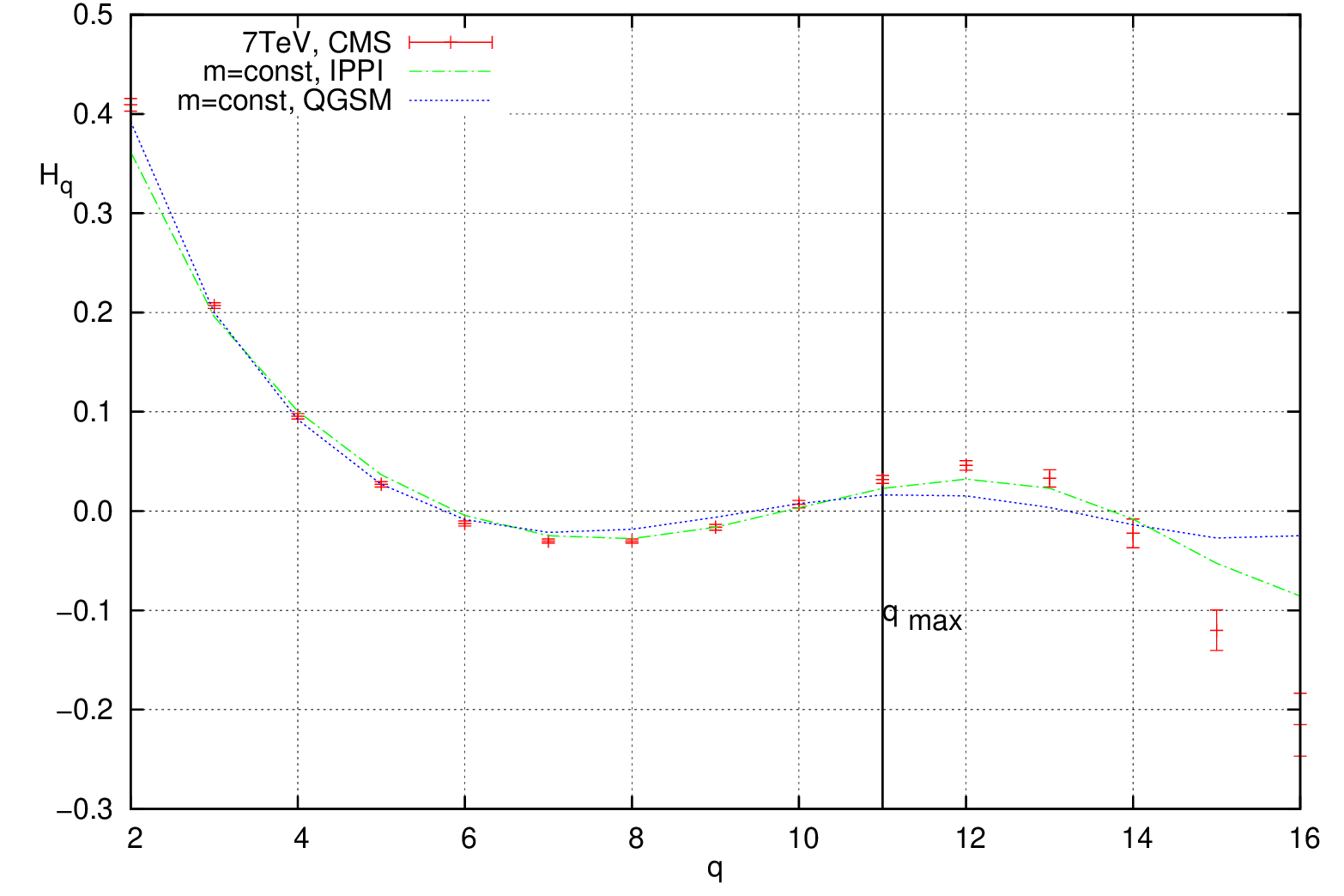}
 \caption{}
 \label{fitHmq7000}
\end{figure}

Afterwards we don't consider the moments starting from such rank $q$ that
satisfies the condition
\begin{equation}
\frac{\ErrorH_q-\langle \ErrorH\rangle_q}{\sqrt{D^H_q}}>1.
\end{equation}
And in any case we don't consider ranks larger than 16. The maximum number of 
the pair parton collisions $j_{max}$ is then chosen automatically so
that to minimize the value in Eq. (\ref{fitPmq=const}). In that respect
our procedure improves the approach of \cite{dnec}, where this parameter
was chosen just among two nearest possibilities. After that one can apply this
procedure in attempts to fit experimental data. At 7 TeV we introduce  the cut
at $q_{max}$=11 (see Fig.\ref{fitHmq7000}). Namely these results are
described in the main content of the paper.

\section{Appendix 2}

\begin{figure}[h]
%fitPH1800.gnu
 \includegraphics[width=\textwidth]{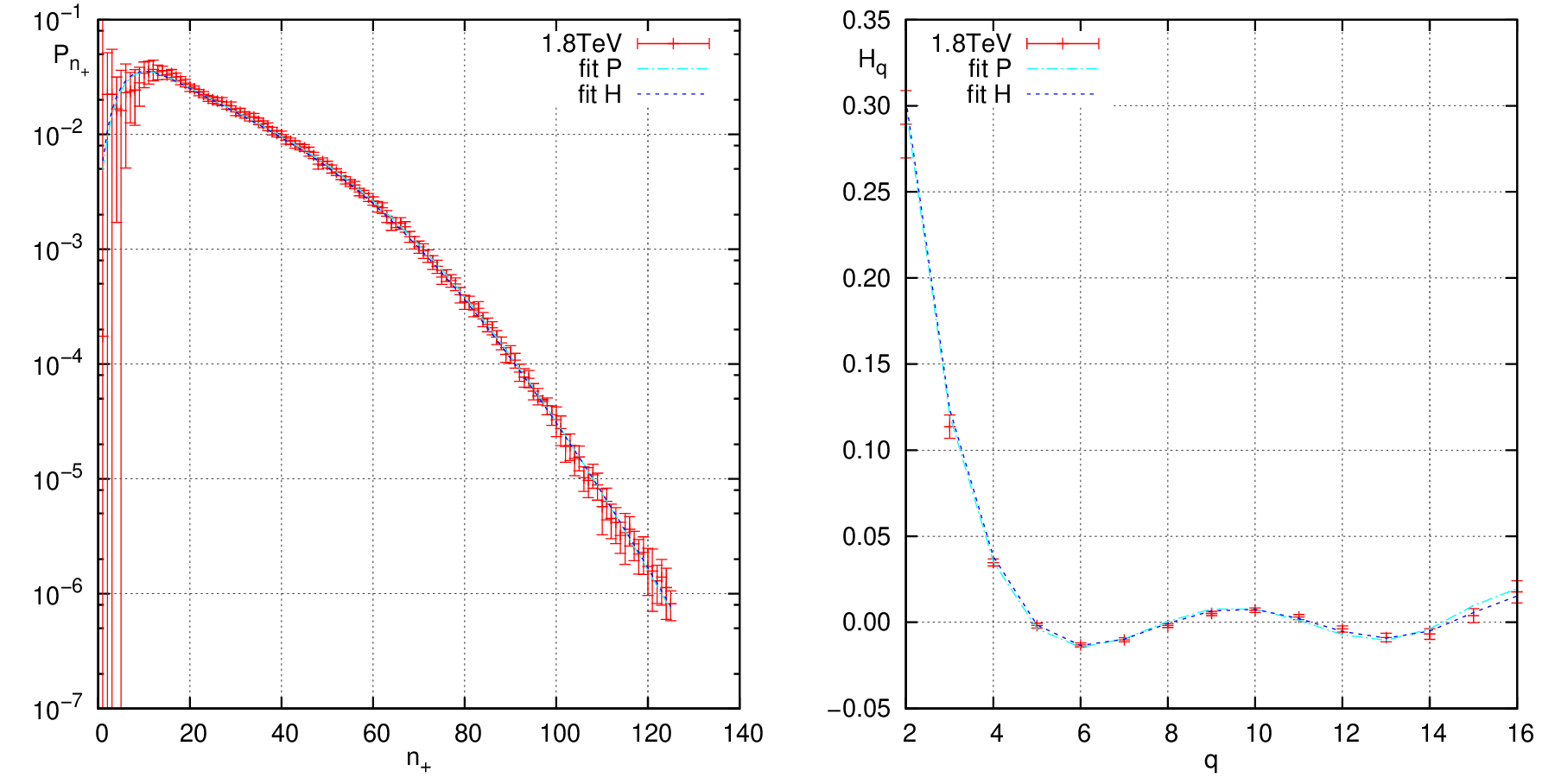}
 \caption{}
 \label{fitPH1800}
\end{figure}

\begin{figure}[h]
%m_q.gnu
 \includegraphics[width=\textwidth]{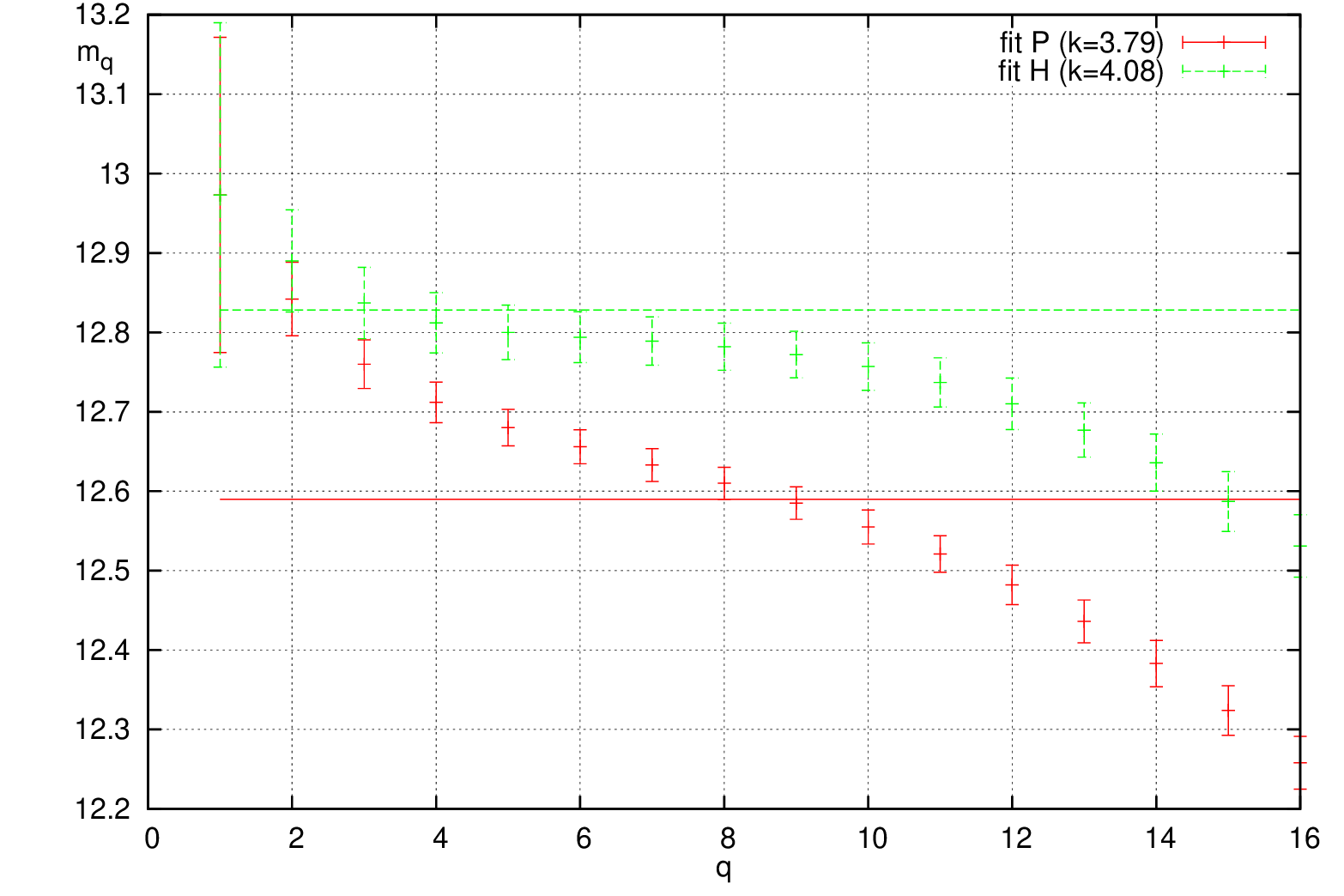}
 \caption{}
 \label{mq1800}
\end{figure}

The $q$-independence of $m$ is a crucial test of the quality of fits.
Let us just briefly mention that we tried to apply the minimization procedure
without any condition of constancy of $m$, i.e. with no last terms in
Eqs. (\ref{fitPmq=const}) and (\ref{fitHmq=const}). It showed slightly different 
(albeit within the limits of less than 10 per cents) values of parameters $k$ 
and $m$ if the Tevatron data at 1.8 TeV were used. Visually, these fits looked 
quite satisfactory (see Fig.~\ref{fitPH1800}) both for multiplicity 
distributions and for $H_q$. However the conditon $m(q,k)$=const is not well 
fulfilled. Minimization of probabilities lead to values $k$=3.79 and $m$=12.59, 
while minimization of $H_q$ gives $k$=4.08 and $m$=12.828. Fig.~\ref{mq1800} 
demonstrates the $q$-dependence of $m(q,k)$ (\ref{frac}) for $k$=3.79 and 
$k$=4.08 (the straight lines show the average values). It is much more 
noticeble than that in Fig.~\ref{fit_mq1800}. Therefore the last fit discussed 
in the paper is preferable.

\medskip
{\bf Acknowledgments}
\medskip

This work was supported by RFBR grants 09-02-00741, 12-02-91504 and by the 
RAS-CERN program.

\end{document}